\documentstyle[12pt]{article}

\begin{document}

\author{E.Sh.Mamasakhlisov \& V.F.Morozov}
\title{{\LARGE Dynamic\ Properties of\ Random\ Heteropolymer\ with\
Correlations
in\ Sequence.}}
\date{30.09.98}
\maketitle

\newpage

\begin{center}
{\bf Abstract}
\end{center}

The some dynamic properties of a random heteropolymer in the condensed
state
are studied in the mode coupling approximation. In agreement with recent
report a dynamic friction increasing is predicted for the random
heteropolymer with power-law correlations in comparison with exponential
correlations. In the case of strong power-law correlations ($\beta >\beta
^{*}$) the dynamic friction function diverge in the thermodynamic limit.
The
qualitative explanation is given for the monomer's dynamics slowing down
and
diverged energetic barrier between the frozen and random coil states. The
possible relations with protein's function and evolution are discussed.

\newpage

\section{Introduction.}

From the molecular point of view the prebiotic evolution is the creation
and
subsequent change of the set of sequences of proteins and nucleic acids
from
the primordial soup. The one of important features of biologically
functional proteins is the ability of recognizing an a priori given target
molecule. One of possible mechanisms of such properties origin was
suggested
in [1,2]. It is the fast polymerization of monomers in presence the target
molecule. It was shown that this procedure produced the slight power-law
correlations in sequence of the forming macromolecule due by the
interactions between target molecule and monomers. It was shown also that
similar type of correlations observed in sequences of real proteins. This
result was obtained [1] by averaging over the ensemble of different
sequences of the homological proteins. At the last years the some
interesting ideas were suggested that try to explain the possible ways to
genetic code origin [3,4]. For example, a correspondence between
nucleotide
and amino acid coding units, may have origin from the sequence-specific
interactions between abiotically synthesized polynucleotides and
polypeptides [4].This point of view is in interplay with mechanism,
proposed
by Pande et al.[1].

The role that protein's and random heteropolymer's sequence correlation
properties play in their structure and function was investigated
previously
in [5,6]. In particularly, it was shown [6] that in random heteropolymers
(
RHP ) the power-law correlations produced the energetic barrier between
folded and random coil states. This result was interpret as an evidence of
power-law correlations only in non-coding sequences of DNA, that is
corresponded to results of Peng et al. [7,8 ].

Purely dynamic theories in mode coupling fashion were derived previously
in
[9-12] for completely random heteropolymers. . So, Roan and Shakhnovich
[10]
derive a mode-mode coupling theory of the RHP and explicitly solved it for
a
polymer in a good solvent. Thirumalai et al. [11] derived dynamic
equations
in the framework of dynamic functional integral ( DFI ) method for a
Garel,
Leibler and Orland model [13]. The objective of the present study is
random
heteropolymer's dynamics treatment in presence of sequence power-law
correlations. We have considered here the Langevin dynamics of RHP with
correlated sequence in mode-coupling approximation. The main goal of the
present article is the investigation of some dynamic properties of the
condensed state of RHP.

\section{Model and Formalism.}

Let us consider the heteropolymer chain with a frozen sequence of
different
monomers. On the microscopic scale thermal motion in macromolecule
produces
small deformations of valent bonds and angles and causes rotations around
single bonds. It is widely believed [14] that at the low temperatures in
the
long macromolecules the long-wave fluctuations are dominant. Thus, the
global properties of the polymeric chain defined by collisions not merely
between monomers, but rather between parts of the chain. It was shown [15]
that interactions in the polymeric chain may be qualitatively represented
as
interactions in the cloud of ''quasimonomers'' with the renormalized
characteristics of interaction.

Thus, we can assume that the state of the polymer chain described by
coarse-grained beads coordinated $\overrightarrow{r}(s)$ and their momenta
$%
\overrightarrow{p}(s)$ and also by the those coordinates
$\overrightarrow{r}%
_\mu $ and $\overrightarrow{p}_\mu $ that describe the precise local
structure of a polymer.

Specifically, we examine the monomer interaction energy of the form 
\begin{equation}
H=E+K+H_{mic}  \label{1}
\end{equation}
where $K$ is the kinetic energy. The $U_{mic}$ is the potential energy for
interactions of microscopic coordinates with other degrees of freedom.
$E(r)$
is the coarse-grained ''beads-on the string'' Hamiltonian

\begin{equation}
E(r)=\frac{Td}{2a^2}\sum_{s=1}^N(r_s-r_{s+1})^2+\sum_{s<s^{^{\prime
}}}B_{ss^{^{\prime }}}\delta (r_s-r_{s^{^{\prime }}})+V(r)  \label{2}
\end{equation}
where $B_{ss^{^{\prime }}}=B\sigma _s\sigma _{s^{^{\prime }}}$ is second
virial coefficient describing two-particle interactions. $\left\{ \sigma
_s\right\} $ are variables, describing species of monomers. $V(r)$ is the
excluded volume interactions approximation by effective confinement term 
\begin{equation}
V(r)=B^{*}\sum_sr_s^2  \label{3}
\end{equation}
where the constant $B^{*}$chosen so that the radius of gyration becomes
the
value, required by the monomers packing density in the globular state
[16]. $%
B_{ss^{^{\prime }}}$ is the effective second virial coefficient of the
interaction between parts of the chain. $B_{ss^{^{\prime }}}$ represents
the
collective property of all the monomers of the chain. Taking into account
the central limit theorem, we can choose $\left\{ \sigma _s\right\} $ as
random variables, characterized by Gaussian distribution function in the
form 
\begin{equation}
P\left\{ \sigma _s\right\} \propto \exp \left[ -\frac 12\left( 
\overrightarrow{\sigma },\widehat{K}^{-1}\overrightarrow{\sigma }\right)
\right]  \label{4}
\end{equation}
where $\overrightarrow{\sigma }=\left\{ \sigma _s\right\} $. $\widehat{K}$
is the matrix describing correlations in the chain sequence 
\begin{equation}
K(\sigma _s,\sigma _{s^{^{\prime }}})=\overline{\sigma _s\sigma
_{s^{^{\prime }}}}  \label{5}
\end{equation}

Following the generalized Rause theory, derived by Hess [17,18] for the
entangled polymeric liquids, consider the dynamic equations, described
random heteropolymer behavior.

Let us a complete set of degrees of freedom is $\left\{ X_i\left( t\right)
\right\} $. Then the equation of motion for $X_i\left( t\right) $ is 
\begin{equation}
\frac d{dt}X_i\left( t\right) =-\widehat{L}X_i(t)  \label{6}
\end{equation}
where $\widehat{L}$ is Liouvillian given as 
\begin{equation}
\widehat{L}=-\sum_{s=1}^N\left[ v(s)\nabla _{r(s)}-\frac 1m\nabla
_{r(s)}H\nabla _{v(s)}\right] -\sum_\mu (v_\mu \nabla _\mu -\frac 1{m_\mu
}\nabla _{r_\mu }H_{mic}\nabla _{v_\mu })  \label{7}
\end{equation}
$v(s)$ and $v_\mu $ are velocities of beads ( segments )and of microscopic
constituents. $m$ and $m_\mu $ are the corresponding masses. The
projection
operator formalism shows that Liouville equation can be reduced into the
form of generalized Langevin equation [17,18,19] 
\begin{equation}
\frac d{dt}r(s,t)=-T\sum_{s^{^{\prime }}s^{^{\prime \prime
}}}\int\limits_0^tdt^{^{\prime }}\mu _{ss^{^{\prime }}}(t-t^{^{\prime
}})\gamma _{s^{^{\prime }}s^{^{\prime \prime }}}r(s^{^{\prime \prime
}},t^{^{\prime }})+g(s,t)  \label{8}
\end{equation}
where $g(s,t)$ satisfied to fluctuation dissipation theorem 
\begin{equation}
\left\langle g(s,t)g(s^{^{\prime }},t^{^{\prime }})\right\rangle =T\mu
_{ss^{^{\prime }}}(t-t^{^{\prime }})  \label{9}
\end{equation}
$\gamma _{s^{^{\prime }}s^{^{\prime \prime }}}$ are the elements of the
nearest-neighbor matrix for an open ideal chain 
\begin{equation}
\gamma _{s^{^{\prime }}s^{^{\prime \prime }}}=-\frac d{a^2}\widehat{I}%
a_{ss^{^{\prime }}}  \label{10}
\end{equation}
As it was shown in [18] the dynamic mobility function fullfil the
following
equation 
\begin{equation}
\mu _{ss^{^{\prime }}}(t)=\frac{2\delta (t)}{\zeta _0}-\frac 1{\zeta
_0}\int\limits_0^td\tau \sum_{s^{^{\prime \prime }}}\mu _{ss^{^{\prime
\prime }}}(t-\tau )\Delta \zeta _{s^{^{\prime \prime }}s^{^{\prime
}}}(\tau )
\label{11}
\end{equation}
where $\Delta \zeta _{s^{^{\prime \prime }}s^{^{\prime }}}(\tau )$ is the
dynamic friction function produced by force correlations between beads (
coarse-grained coordinates ) 
\begin{equation}
\Delta \zeta _{ss^{^{\prime }}}(t)=\frac 1T\langle f_p(s)e^{-\widehat{Q}%
\widehat{L}t}f_p(s^{^{\prime }})\rangle  \label{12}
\end{equation}
\[
\overrightarrow{f}_p(s)=-\nabla _{r(s)}H_{int} 
\]

where $H_{int}=\sum_{s<s^{^{\prime }}}B_{ss^{^{\prime }}}\delta
(r_s-r_{s^{^{\prime }}})$

It was shown that [18] 
\begin{eqnarray}
\Delta \zeta _{ss^{^{\prime }}}(t) &=&\beta \langle \overrightarrow{f}%
_p(s)e^{-\widehat{Q}\widehat{L}t}\overrightarrow{f}_p(s^{^{\prime
}})\rangle
=\beta \langle \overrightarrow{f}_p(s)e^{-\widehat{L}t}\overrightarrow{f}%
_p(s^{^{\prime }})\rangle +  \label{13} \\
&&\ \ \ \ +\beta \sum_{s^{^{\prime \prime }}=1}^N\int\limits_0^td\tau
\langle \overrightarrow{f}_p(s)e^{-\widehat{L(}t-\tau
)}\overrightarrow{v}%
(s^{^{\prime \prime }})\rangle \Delta \zeta _{s^{^{\prime \prime
}}s^{^{\prime }}}(\tau )  \nonumber
\end{eqnarray}
The two-particle interaction forces given as 
\begin{equation}
\overrightarrow{f}_p(s)=-\nabla _{r(s)}H_{int}=-\left| b\right| T\int_qi%
\overrightarrow{q}e^{-i\overrightarrow{q}\overrightarrow{r}(s)}\sigma
_sm(-%
\overrightarrow{q})  \label{14}
\end{equation}
where 
\begin{equation}
m(\overrightarrow{q})=\sum_{s^{^{\prime }}}m_{s^{^{\prime }}}(%
\overrightarrow{q})=\sum_{s^{^{\prime }}}\sigma _{s^{^{\prime }}}\exp
\left[
-i\overrightarrow{q}\overrightarrow{r}(s^{^{\prime }})\right]  \label{15}
\end{equation}
and 
\[
\int_q...=\int \frac{d^dq}{(2\pi )^d} 
\]
are defined by analogy with the bead's collective density fluctuations.
Thus, the dynamic friction function is equal to 
\begin{eqnarray*}
\Delta \zeta _{ss^{^{\prime }}}(t) &=&-\left| b\right| ^2T\int_q\int_p%
\overrightarrow{q}\overrightarrow{p}\sum_{s_1s_2}\langle
m_s(\overrightarrow{%
q})m_{s_1}(-\overrightarrow{q})e^{-\widehat{L}t}m_{s^{^{\prime }}}(%
\overrightarrow{p})m_{s_2}(-\overrightarrow{p})\rangle - \\
&&\ \ \ -\left| b\right| \sum_{s^{^{\prime \prime }}}\int_q\int^td\tau
\sum_{s_1}i\overrightarrow{q}\langle m_s(\overrightarrow{q})m_{s_1}(-%
\overrightarrow{q})e^{-\overrightarrow{L}(t-\tau )}\overrightarrow{v}%
(s^{^{\prime \prime }})\rangle \Delta \zeta _{s^{^{\prime \prime
}}s^{^{\prime }}}(t)
\end{eqnarray*}
The four-point correlation function at the right side of the last equation
can not be calculated rigorously. Following [18] choose as the simplest
approximation for collective density fluctuations 
\begin{equation}
m_s(\overrightarrow{q},t)=\sum_{s^{^{\prime }}}R_{ss^{^{\prime }}}(%
\overrightarrow{q},t)m_{s^{^{\prime }}}(\overrightarrow{q},0)\equiv
\sum_{s^{^{\prime }}}R_{ss^{^{\prime
}}}(\overrightarrow{q},t)m_{s^{^{\prime
}}}(\overrightarrow{q})  \label{16}
\end{equation}
The relaxation functions $R_{ss^{^{\prime }}}(\overrightarrow{q},t)$ are
defined by expression 
\[
\langle m_s(\overrightarrow{q},t)m_{s^{^{\prime }}}(-\overrightarrow{q}%
)\rangle =\sum_{s^{^{\prime \prime }}}R_{ss^{^{\prime \prime }}}(%
\overrightarrow{q},t)\langle m_{s^{^{\prime \prime }}}(\overrightarrow{q}%
)m_{s^{^{\prime }}}(-\overrightarrow{q})\rangle 
\]
In this approximation 
\begin{eqnarray*}
\Delta \zeta _{ss^{^{\prime }}}(t) &=&-\left| b\right| ^2T\int_q\int_p%
\overrightarrow{q}\overrightarrow{p}\sum_{s_1s_2}\sum_{s^{^{\prime \prime
}}s_3}R_{s^{^{\prime }}s^{^{\prime \prime }}}(\overrightarrow{p}%
,t)R_{s_2s_3}(-\overrightarrow{p},t) \\
&&\ \langle m_s(\overrightarrow{q})m_{s_1}(-\overrightarrow{q}%
)m_{s^{^{^{\prime \prime
}}}}(\overrightarrow{p})m_{s_3}(-\overrightarrow{p}%
)\rangle - \\
&&\ \ \ \ -\left| b\right| \int_q\int^td\tau (i\overrightarrow{q}%
)\sum_{s_1}\langle
m_s(\overrightarrow{q})m_{s_1}(-\overrightarrow{q})e^{-%
\widehat{L}t}v(s^{^{\prime \prime }})\rangle \Delta \zeta _{s^{^{\prime
\prime }}s^{^{\prime }}}(\tau )
\end{eqnarray*}
It was shown that [18] 
\begin{eqnarray}
\Delta \zeta _{ss^{^{\prime }}}(t) &=&-\left| b\right| ^2T\int_q\int_p%
\overrightarrow{q}\overrightarrow{p}\sum_{s_1s_2}\sum_{s^{^{\prime \prime
}}s_3}R_{s^{^{\prime }}s^{^{\prime \prime }}}(\overrightarrow{p}%
,t)R_{s_2s_3}(-\overrightarrow{p},t)  \nonumber  \label{17} \\
&&\langle m_s(\overrightarrow{q})m_{s_1}(-\overrightarrow{q}%
)m_{s^{^{^{\prime \prime
}}}}(\overrightarrow{p})m_{s_3}(-\overrightarrow{p}%
)\rangle
\end{eqnarray}
At the integration over $\overrightarrow{q}$ the expression 
$$
\langle m_s(%
\overrightarrow{q})m_{s_1}(-\overrightarrow{q})m_{s^{^{^{\prime \prime
}}}}(%
\overrightarrow{p})m_{s_3}(-\overrightarrow{p})\rangle 
$$ 
gives nonzero
contribution only when beads $s_1$ and $s$ are in contact. However, the
contact between $s_1$ and $s$ have influence on the contact between the
beads $s_3$ and $s^{^{\prime \prime }}$, if they belongs to subchain $%
(s_1-s) $ ( see fig.1 ). The loop formation between $s_1$ and $s$
sufficiently reduced the available conformations space as 
\begin{equation}
\frac{\left\{ {\rm Number\, of\, subchain}(s_1-s){ \rm \, loop\,
conformations\,}%
\right\} }{\left\{ {\rm \, Total\, number\, of\, subchain\, }(s_1-s){\rm
\, conformations\, }\right\} }\propto \left| s_1-s\right|
^{-\frac{d^{*}}2}
\label{18}
\end{equation}
(In the case of ideal Gaussian chain $d^{*}=d$ ).

By this reason, the loop formation between $s_1$ and $s$ have increased
the
probability of contact formation for the beads belongs to subchain
$(s_1-s)$%
. Taking into account this approximation 
\[
\Delta \zeta _{ss^{^{\prime }}}(t)=-\left| b\right| ^2T\int_q\int_p%
\overrightarrow{q}\overrightarrow{p}\sum_{s_1s_2}\sum_{s^{^{\prime \prime
}}s_3\in \left[ s,s_1\right] }R_{s^{^{\prime }}s^{^{\prime \prime }}}(%
\overrightarrow{p},t)R_{s_2s_3}(-\overrightarrow{p},t)\times 
\]
\[
\langle \sigma _s\sigma _{s_1}\sigma _{s^{^{\prime \prime }}}\sigma
_{s_3}e^{-i\overrightarrow{q}\overrightarrow{l}(ss_1)}e^{-i\overrightarrow{p}%
\overrightarrow{l}(s^{^{\prime \prime }}s_3)}\rangle =-\left| b\right|
^2T\int_p\sum_{s_1s_2}\sum_{s^{^{\prime \prime }}s_3\in \left[
s,s_1\right]
}R_{s^{^{\prime }}s^{^{\prime \prime
}}}(\overrightarrow{p},t)R_{s_2s_3}(-%
\overrightarrow{p},t)\times 
\]
\[
\sigma _s\sigma _{s_1}\sigma _{s^{^{\prime \prime }}}\sigma _{s_3}\langle
\int_q\overrightarrow{q}\overrightarrow{p}e^{-i(\overrightarrow{q}+%
\overrightarrow{p})\overrightarrow{l}(ss_1)}\rangle 
\]
where 
\[
\overrightarrow{l}(ss_1)=\overrightarrow{r}(s)-\overrightarrow{r}(s_1) 
\]
The last integral can be evaluated as [18] 
\[
\langle
\int_q\overrightarrow{q}\overrightarrow{p}e^{-i(\overrightarrow{q}+%
\overrightarrow{p})\overrightarrow{l}(ss_1)}\rangle \simeq
-p^2\int_Q\langle
e^{-i\overrightarrow{Q}\overrightarrow{l}(ss_1)}\rangle 
\]
Thus, 
\begin{eqnarray}
\Delta \zeta _{ss^{^{\prime }}}(t) &\simeq &\left| b\right|
^2T\int_p\sum_{s_1s_2}\sum_{s^{^{\prime \prime }}s_3\in \left[
s,s_1\right]
}\sigma _s\sigma _{s_1}\sigma _{s^{^{\prime \prime }}}\sigma
_{s_3}\int_Q\langle
e^{-i\overrightarrow{Q}\overrightarrow{l}(ss_1)}\rangle 
\nonumber  \label{20} \\
&&\ \int_pp^2R_{s^{^{\prime }}s^{^{\prime \prime }}}(\overrightarrow{p}%
,t)R_{s_2s_3}(-\overrightarrow{p},t)
\end{eqnarray}
Because of the integral over $\overrightarrow{p}$ is dominated by the
large $%
p$ modes only correlations between small numbers of beads have
contribution.
Then we assume that in the limit of large wave vectors the correlation
between different beads vanishes and only single bead propagator survives 
\[
\int_pp^2R_{s^{^{\prime }}s^{^{\prime \prime }}}(\overrightarrow{p}%
,t)R_{s_2s_3}(-\overrightarrow{p},t)\approx \delta _{s^{^{\prime
}}s^{^{\prime \prime }}}\delta _{s_2s_3}\int_pp^2G^2(\overrightarrow{p},t) 
\]
where $G(\overrightarrow{p},t)$ is the single bead propagator.

Consequently, dynamic friction function is 
\begin{equation}
\Delta \zeta _{ss^{^{\prime }}}(t)=\left| b\right| ^2T\Psi _{ss^{^{\prime
}}}(\overrightarrow{\sigma })\int_pp^2G^2(\overrightarrow{p},t)
\label{21}
\end{equation}
where 
\[
\Psi _{ss^{^{\prime }}}(\overrightarrow{\sigma })=\sum_{s_1\geq
s^{^{\prime
}}}\sum_{s_2\in \left[ s,s_1\right] }\sigma _s\sigma _{s_1}\sigma
_{s^{^{\prime }}}\sigma _{s_2}\int_Q\langle e^{-i\overrightarrow{Q}%
\overrightarrow{l}(ss_1)}\rangle 
\]
Thus, the generalized Langevin equation for the beads coordinates 
\begin{equation}
\frac d{dt}r(s,t)=-\frac{Td}{a^2}\sum_{s^{^{\prime }}s^{^{\prime \prime
}}}\int^tdt^{^{\prime }}\mu _{ss^{^{\prime }}}(t-t^{^{\prime
}})A_{s^{^{\prime }}s^{^{\prime \prime }}}r(s^{^{\prime \prime
}},t^{^{\prime }})+g(s,t)  \label{22}
\end{equation}
where 
\[
\langle g(s,t)\rangle =0 
\]
\begin{equation}
\langle g(s,t)g(s^{^{\prime }},t^{^{\prime }})\rangle =T\widehat{1}\mu
_{ss^{^{\prime }}}(t-t^{^{\prime }})  \label{23}
\end{equation}
\[
\mu _{ss^{^{\prime }}}(t)=\frac{2\delta (t)\delta _{ss^{^{\prime
}}}}{\zeta
_0}-\frac 1{\zeta _0}\int\limits_0^td\tau \sum_{s^{^{\prime \prime }}}\mu
_{ss^{^{\prime \prime }}}(t-\tau )\Delta \zeta _{s^{^{\prime \prime
}}s^{^{\prime }}}(\tau ) 
\]
The dynamic friction functions $\Delta \zeta _{ss^{^{\prime }}}(t)$ are
defined by eqn.(20). In spite of the case investigated in [18], the $%
A_{ss^{^{\prime }}}$ are the elements of the nearest-neighbor matrix for
the
open chain with the confinement interactions (3).

Let us estimate the magnitude of the dynamic friction function in
dependence
of correlations in $\overrightarrow{\sigma }=\left\{ \sigma _1,...,\sigma
_N\right\} $. Value of the $\Psi _{ss^{^{\prime }}}(\overrightarrow{\sigma
}%
) $ is defined by the summation over large number of beads and formally
similar to the energy with corresponding interaction potential. Thus,
$\Psi
_{ss^{^{\prime }}}(\overrightarrow{\sigma })$ is the extensive quantity.
It's reasonable to suppose that $\Psi _{ss^{^{\prime }}}(\overrightarrow{%
\sigma })$ is the self-averaging quantity. In this case 
\begin{equation}
\Psi _{ss^{^{\prime }}}(\overrightarrow{\sigma })=\overline{\Psi
_{ss^{^{\prime }}}(\overrightarrow{\sigma })}=\sum_{s_1\geq s^{^{\prime
}}}\sum_{s_2\in \left[ s,s_1\right] }\overline{\sigma _s\sigma
_{s_1}\sigma
_{s^{^{\prime }}}\sigma _{s_2}\int_Q\langle e^{-i\overrightarrow{Q}%
\overrightarrow{l}(ss_1)}\rangle }  \label{24}
\end{equation}
The main object of our interests is the condensed state of random
heteropolymer. It is well known that in this state any subchain behaves as
ideal and one-time ( equilibrium ) correlations described by Gaussian
statistics. The deviation from the Gaussian statistics produced a great
entropy lost by order $\simeq N$. Consequently, 
\[
\langle e^{-i\overrightarrow{Q}\overrightarrow{l}(ss_1)}\rangle
=e^{-Q^2b^2\left| s-s_1\right| } 
\]
where $b$ is the length of the effective statistical segment. Thus, 
\[
\Psi _{ss^{^{\prime }}}(\overrightarrow{\sigma })=\int_Q\Phi
_{ss^{^{\prime
}}}(\overrightarrow{Q}) 
\]
where 
\begin{eqnarray}
\Phi _{ss^{^{\prime }}}(\overrightarrow{Q}) &=&\sum_{s_1\geq s^{^{\prime
}}}\sum_{s_2\in \left[ s,s_1\right] }\overline{\sigma _s\sigma
_{s_1}\sigma
_{s^{^{\prime }}}\sigma _{s_2}e^{-Q^2b^2\left| s-s_1\right| }}\approx
\label{25} \\
\ &\approx &\sum_{s_1\geq s^{^{\prime }}}\sum_{s_2\in \left[ s,s_1\right]
}%
\overline{\sigma _s\sigma _{s_1}\sigma _{s^{^{\prime }}}\sigma _{s_2}}%
e^{-Q^2b^2\left| s-s_1\right| }  \nonumber
\end{eqnarray}
Because of variables $\overrightarrow{\sigma }$ are Gaussian distributed
and 
$\overline{\sigma _s}=0$, then 
\[
\overline{\sigma _s\sigma _{s_1}\sigma _{s^{^{\prime }}}\sigma _{s_2}}=%
\overline{\sigma _s\sigma _{s_1}}\cdot \overline{\sigma _{s^{^{\prime
}}}\sigma _{s_2}}+\overline{\sigma _s\sigma _{s^{^{\prime }}}}\cdot 
\overline{\sigma _{s_1}\sigma _{s_2}}+\overline{\sigma _s\sigma
_{s_2}}\cdot 
\overline{\sigma _{s_1}\sigma _{s^{^{\prime }}}} 
\]
$Q$ mode of the $\Psi _{ss^{^{\prime }}}(\sigma )$ may be estimated as 
\begin{eqnarray}
\Phi _{ss^{^{\prime }}}(\overrightarrow{Q}) &=&\sum_{s_1\geq s^{^{\prime
}}}%
\overline{\sigma _s\sigma _{s_1}}e^{-Q^2b^2\left| s-s_1\right|
}\sum_{s_2\in
\left[ s,s_1\right] }\overline{\sigma _{s^{^{\prime }}}\sigma _{s_2}}+
\label{26} \\
&&\ \ \ \ +\sum_{s_1\geq s^{^{\prime }}}\overline{\sigma _s\sigma
_{s^{^{\prime }}}}e^{-Q^2b^2\left| s-s_1\right| }\sum_{s_2\in \left[
s,s_1\right] }\overline{\sigma _{s_1}\sigma _{s_2}}+  \nonumber \\
&&\ \ \ \ +\sum_{s_1\geq s^{^{\prime }}}\overline{\sigma _{s_1}\sigma
_{s^{^{\prime }}}}e^{-Q^2b^2\left| s-s_1\right| }\sum_{s_2\in \left[
s,s_1\right] }\overline{\sigma _s\sigma _{s_2}}  \nonumber
\end{eqnarray}
It is obvious that in dependence of rate of correlation function decay the
magnitude of $\Phi _{ss^{^{\prime }}}(\overrightarrow{Q})$ may be quite
different.

First of all, let us consider the exponential decay of inter monomer's
correlations. In this case 
\begin{equation}
K(s,s^{^{\prime }})\propto \exp \left( -\frac{\left| s-s^{^{\prime
}}\right| 
}\xi \right)  \label{27}
\end{equation}
For simplicity, we have restricted our consideration by the case $s\approx
s^{^{\prime }}$. Then the main term, contributed to $\Phi _{ss^{^{\prime
}}}(%
\overrightarrow{Q})$ is 
\begin{eqnarray*}
\sum_{s_1\geq s^{^{\prime }}}\overline{\sigma _s^2}e^{-Q^2b^2\left|
s-s_1\right| }\sum_{s_2\in \left[ s,s_1\right] }\overline{\sigma
_{s_1}\sigma _{s_2}} &=& \\
\ &=&K\sum_{s_1\geq s^{^{\prime }}}e^{-Q^2b^2\left| s-s_1\right|
}\sum_{s_2\in \left[ s,s_1\right] }\overline{\sigma _{s_1}\sigma _{s_2}}
\end{eqnarray*}
It is obvious that for the exponential correlations 
\[
\sum_{s_2\in \left[ s,s_1\right] }\overline{\sigma _{s_1}\sigma _{s_2}}%
=a_{ss_1}<\infty 
\]
is finite for any range of summation. Consequently, 
\[
\lim_{N\longrightarrow \infty }\Phi _{ss^{^{\prime }}}(\overrightarrow{Q}%
)<\infty 
\]
for any value of $Q$. Moreover, 
\[
\int_Q\sum_{s_1\geq s^{^{\prime }}}e^{-Q^2b^2\left| s-s_1\right|
}\sum_{s_2\in \left[ s,s_1\right] }\overline{\sigma _{s_1}\sigma _{s_2}}%
\simeq b^{-d}a\sum_{s_1\geq s}\left| s-s_1\right| ^{d/2} 
\]
is finite in the thermodynamic limit ( $N\gg 1$ ). In spite of exponential
correlations, the power-law decay leads to the following picture. Suppose
that 
\begin{equation}
K(s,s^{^{\prime }})\propto \left| s-s^{^{\prime }}\right| ^{\beta -1}
\label{28}
\end{equation}
Then, 
\begin{equation}
\Phi _{ss^{^{\prime }}}(\overrightarrow{Q})=K\sum_{s_1\geq
s}e^{-Q^2b^2\left| s-s_1\right| }\sum_{s_2\in \left[ s,s_1\right]
}\overline{%
\sigma _{s_1}\sigma _{s_2}}\simeq
K\sum_{M=0}^Ne^{-Q^2b^2M}\sum_{S=0}^M\left| M-S\right| ^{\beta -1}
\label{29}
\end{equation}
It is known that [20] the last sum may be asymptotically estimated as 
\[
\sum_{S=0}^M\left| M-S\right| ^{\beta -1}\leq Const\left| 1-\beta \right|
^{\beta -1}+\frac{M^\beta }\beta +\frac 1{2M^{1-\beta }} 
\]
for $M\gg 1$. Consequently, 
\[
\Phi _{ss^{^{\prime }}}(\overrightarrow{Q})\simeq K\sum_{M=0}^NM^\beta
e^{-Q^2b^2M} 
\]
Thus, 
\begin{equation}
\Psi _{ss^{^{\prime }}}(\overrightarrow{\sigma })=\int_Q\Phi
_{ss^{^{\prime
}}}(\overrightarrow{Q})\simeq b^{-d}K\sum_{M=0}^NM^{\beta -d/2}
\label{30}
\end{equation}
In the thermodynamic limit $\Psi _{ss^{^{\prime }}}(\overrightarrow{\sigma
}%
) $ diverge if 
\begin{equation}
\beta >\beta ^{*}=d/2-1  \label{31}
\end{equation}
In the case of three-dimensional space the ''critical value'' of $\beta $
is 
\begin{equation}
\beta ^{*}=\frac 12  \label{32}
\end{equation}
Thus, the magnitude of dynamic friction function is strictly dependent
from
the type of correlations. For the exponential decay the dynamic mobility
is
more much than in the case of log-range ( power-law ) correlations.
Moreover, for the strong ($\beta >\beta ^{*}$) power-law correlations the
long heteropolymer chain ($N\rightarrow \infty $) exhibits a zero mobility
in the globular state.

This situation is similar to the case of strong entanglements, described
in
[18], where 
\[
\int\limits_0^td\tau \Delta \zeta (\tau )\simeq N 
\]

\section{Discussion}

The subject of present article is investigation of relation between the
protein's sequence and its dynamics.

In the case of power-law correlations the dynamic friction behavior is
qualitatively distinguished from those in the heteropolymer with
short-range
correlations. While, for the exponential correlations the $\Delta \zeta
_{ss^{^{\prime }}}(t)$ is finite in the thermodynamic limit, in the case
of
power-law correlation $(\beta \geq 1/2)$ 
\begin{equation}
\Delta \zeta _{ss^{^{\prime }}}(t)\propto N^{\beta -1/2}  \label{33}
\end{equation}
$(N\gg 1)$. Consequently, the dynamic mobility of chain segments in the
globular state sufficiently decreased with the range of correlations
increasing.

This result give us an ability to explain the possible mechanism of large
energetic barrier existence between frozen and random coil states that was
shown in [6]. The long-range correlations in sequence must be increased
the
probability of advantageous contacts for the neighbor and especially
non-nearest neighbor monomers of chain that effectively slowing down the
mutual slip of the chain parts ( see fig.2). It may be explain the so
strong
dependence of dynamic mobility from the range of correlations. The
equilibrium properties of the RHP defined by averaging over the all chain
conformations. For the long-range correlations in the chain sequence the
extended regions of similar monomers are dominated. In the globular state,
characterized by monomer's position fluctuations, the averaging over all
conformations included also a very disadvantageous contacts between
extended
regions of different monomers. This produce a great energetic barrier for
the RHP with power-law correlations.

What may be the constructive role of such correlations in sequence?

One of the necessary properties of biologically functional protein is it
ground state stability, provided by the energetic barrier existence
between
native and misfolded states. The long heteropolymer chain have a very low
mobility in the state of disordered globule and denaturation/renaturations
processes are very difficult in this state. But the real protein's single
chain length in most cases don't exceed $\simeq 10^2$ amino acids. It is
due
not only by the possible errors of replication from the very long segments
of RNA. It may be related with necessity to have the stable native state.
It
is obvious that the random coil may be transferred to the folded ( native
)
state only through the disordered globule state. It's due by the low
entropy
of the Gaussian chains in this state in comparison with those in any other
state of the polymeric chain [14]. The decreased mobility of disordered
globule state should be stabilize the native state of the system.

The results, described above may be relevant to those, obtained by Pande
et
al. [1,2]. The power-law correlations in sequence can additionally
stabilized the frozen state of the heteropolymeric globule.

\newpage

{\bf Figure 1.}

The contact formation between the chain residues $s$ and $s_1$ increased
the
probability of the internal residues ($s^{^{\prime \prime }}$ and $s_3$)
contact formation.

{\bf Figure 2.}

Schematic representation of the mutual movement of the chain parts in the
presence of the long-range correlations. ${\bf z}${\bf \ }are the
disadvantageouse contacts between different monomers and ${\bf x}$ are the
advantageouse contacts between similar monomers.

\end{document}